\documentclass[twocolumn,aps,prb,amssymb,superscriptaddress]{revtex4-2}

\usepackage{graphicx}
\usepackage{lipsum}
\usepackage{float}
\usepackage{amsmath}
\usepackage{mathtools}
\usepackage{bbold}
\usepackage{physics}
\usepackage{ulem}
\usepackage{siunitx}
\usepackage[dvipsnames]{xcolor}

\usepackage[utf8]{inputenc}

\usepackage[breaklinks=true]{hyperref}
\hypersetup{
        colorlinks = true,
        urlcolor = blue,
        citecolor = blue
}

\begin{document}

\title{Electron-optics using negative refraction \\in two-dimensional inverted-band $pn$ junctions} 

\author{Yuhao Zhao}
\email{yzhao.phy@gmail.com}
\affiliation{\mbox{Institute for Theoretical Physics, ETH Zurich, 8093 Zurich, Switzerland}}
\author{Anina Leuch}
\affiliation{\mbox{Institute for Theoretical Physics, ETH Zurich, 8093 Zurich, Switzerland}}
\author{Oded Zilberberg}
\affiliation{\mbox{Department of Physics, University of Konstanz, D-78457 Konstanz, Germany}}
\author{Antonio \v{S}trkalj}
\email{astrkalj@phy.hr}
\affiliation{\mbox{T.C.M. Group, Cavendish Laboratory,
    University of Cambridge, Cambridge, CB3 0HE, United Kingdom}}
\affiliation{\mbox{Department of Physics, Faculty of Science, University of Zagreb, Bijenička c. 32, 10000 Zagreb, Croatia}}

\begin{abstract}
  Electron optics deals with condensed matter platforms for manipulating and guiding electron beams with high efficiency and robustness. Common devices rely on the spatial confinement of the electrons into one-dimensional (1D) channels.  Recently, there is growing interest in electron optics applications in two dimensions, which heretofore are almost exclusively based on graphene devices.
  In this work, we study band-inverted systems resulting from particle-hole hybridization and demonstrate their potential for electron optics applications. We develop the theory of interface scattering in an inverted-band $pn$ junction using a scattering matrix formalism and observe negative refraction conditions as well as transmission filtering akin to graphene's Klein tunneling but at finite angles. Based on these findings, we provide a comprehensive protocol for constructing electron optic components, such as focusing and bifurcating lenses, polarizers, and mirrors. We numerically test the robustness of our designs to disorder and finite temperatures, and motivate the feasibility of experimental realization. Our work opens avenues for electron optics in two dimensions beyond graphene-based devices, where a plethora of inverted-band materials in contemporary experiments can be harnessed. 
\end{abstract}
\maketitle

%%%%%%%%%%%%%%%%%%%%%%%%%%%%%%%%%%%%%%%%%%%%%%%	
%%%%%%%%%%%%%%%%%%%%%%%%%%%%%%%%%%%%%%%%%%%%%%%	
\section{Introduction   \label{sec:Introduction}}
Ballistic electrons in clean two-dimensional electron gas (2DEG) share many similarities with light scattering in classical optics, namely, they demonstrate their wave nature by propagating through the system in ray-like paths and get reflected or transmitted when impinging on an interface. 
This wave-like behavior of electrons in ultra-clean materials inspires studies of electron optics applications in condensed matter systems, which aim to reproduce and enhance existing effects observed with light, e.g. focusing of incident beams, as well as introduce new effects that are not present in classical optical systems, with negative refraction being one of them.
However, unlike in optical systems where it is possible to produce highly concentrated unidirectional beams of light, realization of electron optics is suffering from lack of control over the injected electrons' direction, as well as the destruction of their coherence by scattering with impurities that are naturally present in every solid-state system.
While both effects make it challenging to form a well-defined trajectory for electron beams, the latter further restricts the realization of electron optics devices to the mesoscopic regime, where the trajectory of electrons remains coherent. 
Regarding the control over the directionality, efforts have been made to build one-dimensional (1D) waveguides for electrons in condensed matter systems, e.g., by confining them into quantum wires~\cite{liang2001,das2012a, Annadi2018} or at the edges of quantum Hall devices~\cite{ozyilmaz2007a,amet2014a,nichele2016,nguyen2016a}. 
%%%%%%%%%%%%%%%%%%%%%%%%%%%%%%%%%%%%%%%%%%%%%%%	
\begin{figure}[!ht]
	\centering
	\includegraphics[scale=1]{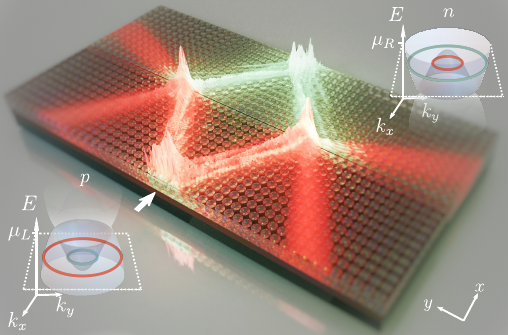}
	\caption{Illustration of a scattering event at an inverted-band $pn$ junction. Due to the inverted-band structure, hole-like spinors (red) injected from a point source (white arrow) in the $p$-doped region scatter at the $pn$ interface and focus as transmitted particle-like spinors (green) in the $n$-doped region. We superimpose an exemplary local density of states of a scattering trajectory of particle-hole spinors, cf.~Eq.~\eqref{eq:LDOP} and Fig.~\ref{Fig:scattering amplitudes for angles and local densities} for the derivation and numerical implementation.
    Insets: The sombrero-hat shaped band structure of the $p$- and $n$-doped regions. The red and green concentric circles mark the hole-like and particle-like branches at the respective Fermi surfaces, i.e., at chemical potentials $\mu_{L,R}$ [cf.~Eq.~\eqref{eq:H continuous}].}
	\label{Fig:setup and band}
\end{figure}
%%%%%%%%%%%%%%%%%%%%%%%%%%%%%%%%%%%%%%%%%%%%%%%	

Guiding electron beams in two-dimensional (2D) systems, without using energy-costly external magnetic fields, remains an unsolved challenge. Moreover, optical elements like lenses and mirrors, with which beams can be controlled and manipulated, have no generic counterparts in the context of various electronic condensed matter systems. While in geometric optics, the design of optical elements relies on the comprehensive depiction of the scattering process at the interface using Snell's law and the Fresnel equation, a generic theory of the electronic scattering process at the interface formed between two materials -- known as a junction -- is missing.

The key constituent of electron-optics devices is a $pn$ junction created by external electrostatic gates. Such $pn$ junctions in certain materials can possess positive/negative effective refractive index for electrons, which was recently discussed as a potential platform for realizing electronic Veselago lens~\cite{veselago1968,Cheianov2007}, beam splitters~\cite{rickhaus2015} and collimators~\cite{park_electron_2008,liu_creating_2017}. Some experimental realizations have been achieved in traditional 2DEG~\cite{spector_electron_1990,smith_low-dimensional_1996} and 2D materials based on graphene~\cite{young_quantum_2009,taychatanapat_electrically_2013,lee_observation_2015,Chen2016}. Furthermore, recently, experiments on hybrid systems of polaritons in artificial honeycomb lattices and hyperbolic materials~\cite{Hu2023,Sternbach2023} reported a direct observation of negative refraction.

For a long time, electron optics in 2D was exclusively reserved for graphene-like materials~\cite{Cheianov2007,rickhaus2015,park_electron_2008,liu_creating_2017,young_quantum_2009,taychatanapat_electrically_2013,lee_observation_2015,Chen2016}. Due to the inherent property of Dirac materials known as the Klein tunneling effect~\cite{Klein1929,Katsnelson2006,Shytov2008,young_quantum_2009}, only waves entering the $pn$ junction at a zero incident angle are perfectly transmitted. Note, however, that the  direct observation of electron optics using Klein tunneling is possible only with $pn$ interfaces that are extremely sharp. 
The effect of negative refraction, which is crucial for realizing elements such as the Veselago lens, is diminished as the $pn$ interface becomes wider~\cite{cheianov_selective_2006}.

In a recent work~\cite{Karalic2020}, some of us discovered an alternative mechanism for realizing optical devices in electronic systems. More precisely, a $pnp$ junction-based Fabry-P\'erot interferometer was realized an inverted-band system. 
Materials containing a band-inversion exist vastly in multiple 2D materials such as InAs/GaSb~\cite{Altarelli1983,Lakrimi1997,Yang1997,Cooper1998}, bilayer graphene~\cite{island_spinorbit-driven_2019}, and other topological materials~\cite{Jiang2020}. 
The band structure, shaped as a “sombrero hat”, provides a larger tunability in steering the electron scattering at the interface. In particular, since the corresponding quasiparticles in the system acquire an additional degree of freedom -- being particle- and hole-like owing to the two bands with opposite effective masses -- one can develop intriguing optical concepts such as bifurcating lenses to electronic systems by referring to being particle/hole-as the counterpart of polarization for electrons. 

This paper demonstrates how to create electron optics application using inverted-band $pn$ junctions. We provide a comprehensive theoretical description of the scattering processes at the $pn$ interface, which can be readily created using the appropriate gating potentials. Starting with a simple model featuring a sombrero-hat dispersion discussed in Sec.~\ref{sec:model}, we develop an effective Snell's law for scattering processes categorized into four regimes in Sec.~\ref{sec:theory of scattering}. The corresponding scattering amplitudes are obtained using a scattering matrix approach. We find that, unlike in graphene, the inverted-band $pn$ junction manifests a Klein tunneling alike effect, that filters transmission at finite incident angles. Such scattering facilitates strong negative refraction for both reflected and transmitted waves. In Sec.~\ref{sec:numerical analysis}, we discuss the possible applications of negative refraction and, in Sec.~\ref{sec:devices}, we provide a protocol for constructing of some basic blocks of electron optics: a mirror, a bifurcating lens, a polarizer, and Veselago and focusing lenses. The robustness of these devices against the influence of thermal fluctuations and disorder is probed and confirmed using numerical simulations presented in Sec.~\ref{sec:temperature and disorder}.

%%%%%%%%%%%%%%%%%%%%%%%%%%%%%%%%%%%%%%%%%%%%%%%%%%%%%%%%%%%
%%%%%%%%%%%%%%%%%%%%%%%%%%%%%%%%%%%%%%%%%%%%%%%%%%%%%%%%%
\section{Model  \label{sec:model}}
We consider a spinless 2D two-band model 
\begin{equation}\label{eq:H continuous}
    \mathcal{H}_{L/R}=\begin{pmatrix}
    \mathcal{M}_0+\mathcal{M}_2\mathbf{k}^2 - \mu_{L/R} & \mathcal{A}_c\\
    \mathcal{A}_c&-\mathcal{M}_0-\mathcal{M}_2\mathbf{k}^2 - \mu_{L/R}
    \end{pmatrix} \, ,
\end{equation}
where $\mathbf{k}$ is a 2D wave vector. The parameter $\mathcal{M}_0$ controls the energy overlap between the bands, and $\pm\mathcal{M}_2$ quantifies the effective positive (negative) mass for the particle- (hole-)like band dispersions. The hybridization between the two bands is determined by  $\mathcal{A}_c$. To model a $pn$ junction (see Fig.~\ref{Fig:setup and band}), we shift the chemical potentials in the left and the right region by $\mu_L$ and $\mu_R$, respectively. Thus, $p$- and $n$-doped regions form, when the Fermi surface cuts through the lower/upper band, respectively, see insets of Fig.~\ref{Fig:setup and band}. 
Diagonalizing the Hamiltonian, we obtain the dispersion $E(\mathbf{k}) = -\mu_{L/R} \pm \sqrt{(\mathcal{M}_0+\mathcal{M}_2 |\mathbf{k}|^2)^2+\mathcal{A}_c^2}$ and the corresponding eigenstates $\vec{\Psi}=(\psi_p,\psi_h)^T$. The two components of the spinor represent the contributions of the particle- and hole-like bands. When $\mathcal{M}_0<0$, resulting dispersion resembles two opposing “sombrero hats”, see insets of Fig.~\ref{Fig:setup and band}. In other words, an inverted-band regime appears, ranging between energies $- \Lambda_{\mu} \leq E \leq -\mathcal{A}_c$ and $\mathcal{A}_c \leq E \leq  \Lambda_{\mu}$, where  $\Lambda_{\mu}=\sqrt{\mathcal{M}_0^2+\mathcal{A}_c^2}$ is the energy at which the inverted-band regime ends.

In Fig.~\ref{Fig:setup and band}, we illustrate the $pn$ junction and band structures of each region. We set the potentials $\mu_{L/R}$ within the inverted-band regime, thus creating a Fermi surface that consists of two branches appearing as two concentric circles, see insets of Fig.~\ref{Fig:setup and band}. In the $p$-doped region, the positive (negative) effective mass indicates that the states located at the inner (outer) branch are particle-(hole-)like. Due to the band inversion, such correspondence is reversed in the $n$-doped part, i.e., the states are particle-(hole)-like at the outer (inner) branch. 

As the eigenstates' spinor is represented by a complex-valued two-vector $\vec{\Psi}=(\psi_p,\psi_h)^T$, we can interpret it as an arrow rotating around the Bloch sphere, where the north (south) pole represents the particle-(hole-)like contribution. We thus denote the spinor as a \textit{band spinor} since it rotates between particle- and hole-like bands. In the following, we refer to its pointing direction as the “polarization” of the band spinor, by analogy with the polarization in traditional optics. For a given band spinor, the polarization is defined as $\vec{\mathcal{P}}=\vec{\Psi}^\dagger\vec{\sigma}\vec{\Psi}$, where $\vec{\sigma}=(\sigma_x, \sigma_y, \sigma_z)^\mathrm{T}$ is a vector of Pauli matrices acting on the band degree of freedom. The state is particle-(hole)-like when $\mathcal{P}_z>0$ ($\mathcal{P}_z<0$).

%%%%%%%%%%%%%%%%%%%%%%%%%%%%%%%%%%%%%%%%%%%%%%%		
\section{Scattering at the interface    \label{sec:theory of scattering}}
When a beam of light impinges upon the interface between two optical media, it splits and propagates along two trajectories: one reflects back to the medium from where it originated, while the other transmits through the interface and enters the next medium. These two scattering processes, i.e. reflection and transmission, are cooperatively determined by properties such as the wavelength and the polarization of the incident light beam, as well as the refractive indices of optical media. In our studied electronic $pn$ junction, a similar scattering process manifests for ballistic electrons -- the ones with a mean free path larger than all other scales in the system~\cite{Ihn2009} -- due to their wavelike nature, see  Figs.~\ref{Fig:setup and band}, \ref{Fig:scattering processes} and~\ref{Fig:scattering angles and velocity map}.

\subsection{Scattering states \label{subsec:scattering angles and the scattering tyes}}
%%%%%%%%%%%%%%%%%%%%%%%%%%%%%%%%%%%%%%%%%%%%%%%	
\begin{figure}[t!]
		\centering
		\includegraphics[scale=1]{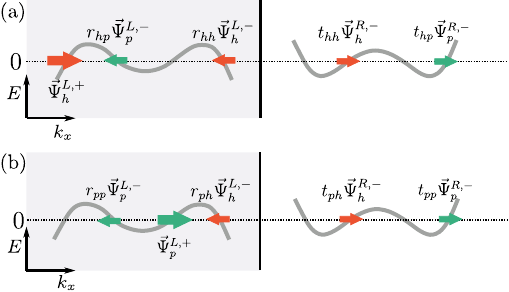}
		\caption{A sketch of the band structure and scattering processes for a one-dimensional channel, i.e., for a specific $k_y$. The dashed horizontal line denotes the Fermi surface at $E=0$. The $k_y$-momentum is chosen such that both regions are in the inverted-band regime. (a) A hole-like spinor $\vec{\Psi}^{L,+}_{h}$ (large red arrow) incidents from the $p$-doped region (gray). (b) A particle-like spinor $\vec{\Psi}^{L,+}_{p}$ (large green arrow) incidents from the $p$-doped region. The reflected $\vec{\Psi}^{L,-}_{a=p/h}$ and transmitted $\vec{\Psi}^{R,-}_{a=p/h}$ states are denoted with their corresponding scattering amplitudes.}
		\label{Fig:scattering processes}
\end{figure}
The scattering processes of electrons in our band-inverted $pn$ junction are more complex than light scattering, as the electrons are redistributed between the band spinors introduced in Sec.~\ref{sec:model}.
The band spinors have a nonlinear energy dispersion, and therefore at constant energy, there can be either one or two incident -- as well as transmitted and reflected -- waves at the $pn$ interface; see Fig.~\ref{Fig:scattering processes}. 
Note that these details lead to various scattering scenarios depending on the chemical potentials of the junctions. On a technical level, this implies that the incident and outgoing band spinors as well as the dominant scattering processes change as a function of chemical potentials. 

All possible coherent scattering processes at the $pn$ interface can be described by the scattering-matrix formalism
\begin{equation}    \label{eq:scattering equation1}
    \begin{pmatrix}
    \vec{\Phi}^{L,-}\\
    \vec{\Phi}^{R,-}
    \end{pmatrix}=
    \mathbf{S}
    \begin{pmatrix}
    \vec{\Phi}^{L,+}\\
    \vec{\Phi}^{R,+}
    \end{pmatrix},
\end{equation}
where $\vec{\Phi}^{l,d}$ is the incoming/outgoing spinor state -- incoming and outgoing concerning the $pn$ interface -- with direction, $d=+/-$, impinging onto the interface from the left/right lead, $l=L/R$. The state $\vec{\Phi}^{l,d}$ denotes the available states for the scattering when assuming energy and momentum conservation. In the most general case, there can be incoming spinor states of both polarizations from the left $p$-doped region, as well as from the right $n$-doped region. Similarly, we have both types of spinors for outgoing states. For simplicity, we restrict our analysis in the following to the case where the states are incoming only from the left region, see processes in Fig.~\ref{Fig:scattering processes}. 

The scattering matrix is given by
\begin{equation}\label{eq:s matrix}
    \mathbf{S}=\begin{pmatrix}
    \mathbf{r} & \mathbf{t}'\\
    \mathbf{t} & \mathbf{r}'
    \end{pmatrix}\,,
\end{equation}
with the scattering amplitudes $\mathbf{r}$, $\mathbf{r}'$, $\mathbf{t}$ and $\mathbf{t}'$ being $2\times2$ matrices. As commented above, since we are only interested in the case where incident waves are impinging from the left, the processes related to $\mathbf{t}'$ and $\mathbf{r}'$ will play no role in the further analysis.
To calculate the relevant scattering amplitudes, we impose the spatial continuity of the states $\vec{\Phi}^{l,d}$ and their derivatives at the $pn$ interface, see Appendix~\ref{app:scattering matrix approach}.
Furthermore, we assume that the $pn$ interface is perfectly sharp, implying the conservation of the perpendicular component of the momentum $\mathbf{k}$, e.g., the conservation of the $\mathbf{k}_\perp\equiv k_y$ momenta for the situation of a scattering in the $x$-direction, depicted in Fig.~\ref{Fig:setup and band}. Crucially, this allows us to treat the 2D scattering problem as a collection of 1D scattering channels labeled by different $k_y$ momenta. 

Before solving Eq.~\eqref{eq:scattering equation1} and obtaining the scattering amplitudes encoded in the matrices $\mathbf{r}$ and $\mathbf{t}$, we first discuss which band spinors contribute to the scattering states $\vec{\Phi}^{l,d}$. If the chemical potential is $|\mu_l|>\Lambda_\mu$, only a single hole-(particle-)like band spinor is present in the $p$-($n$-)doped region. In that case, a hole-like band spinor that propagates towards the $pn$ junction can be transmitted only as a particle-like spinor and reflected solely as a hole-like spinor.
The situation is more involved when the chemical potential is tuned into the inverted-band regime, i.e., when $\mathcal{A}_c < |\mu_l| < \Lambda_\mu$. 
Specifically, as depicted in Fig.~\ref{Fig:scattering processes}, we can have two possibilities for incident band spinors that propagate towards the interface. Depending on the scattering amplitudes, they redistribute into two transmitted and two reflected spinors moving away from the interface. 
In this case, the availability of transmission/reflection channels further depends on the particle's momentum, as the Fermi surface cuts the dispersion and forms two branches, see the insets of Fig.~\ref{Fig:setup and band} as well as Fig.~\ref{Fig:scattering angles and velocity map}(a). The radii of the inner and outer branches are given by 
\begin{equation}
     \Lambda^{l}_{\lessgtr}=\left[-\frac{\mathcal{M}_0}{\mathcal{M}_2}\mp\frac{\sqrt{\mu_{l}^2-\mathcal{A}_c^2}}{\mathcal{M}_2}\right]^{\frac{1}{2}}.
\end{equation}

Hence, for $|k_y|<\Lambda_{<}^{l}$, we have both scattering processes available.
On the other hand, scattering states with $|k_y|>\Lambda_{<}^{l}$ will involve solely hole-(particle)-like spinors in the $p$-($n$)-doped region, respectively. 
Combining the energy and $k_y$ cutoffs, i.e., $\Lambda_{\mu}$ and $ \Lambda^{l}_{\lessgtr}$, respectively, we summarise the four different regimes of the scattering processes in TABLE~\ref{tab:scattering type}. -- according to the polarisation of the propagating spinors.
In each regime, we use, for instance, \textit{ph-p} to denote the case where particle- and hole-like spinors exist in the $p$-doped region and only particle-like spinors exists in the $n$-doped region.
%
%%%%%%%%%%%%%%%%%%%%%%%%%%%%%%%%%%%%%%%%%%%%%%%%%%%%%%%%%%%%%%
\begin{table}
\begin{tabular}{m{0.07\columnwidth}|m{0.45\columnwidth}|m{0.45\columnwidth}}
 ${}_{L}\backslash^{R}$ & \hspace{1.6cm} III &\hspace{1.6cm} IV\\
\hline
 I & \multicolumn{1}{c}{\textit{h-ph}} & \multicolumn{1}{|c}{\textit{h-p}}\\ 
 \hline
 II & \multicolumn{1}{c}{\textit{ph-ph}} & \multicolumn{1}{|c}{\textit{ph-p}}\\ 
\end{tabular}
\caption{\label{tab:scattering type} Scattering regimes for different $|k_y|$ and $|\mu|$ in the left and right regions. In the left, $p$-doped region ($\mu_L<0$), particle- and hole-like spinors coexist when $|k_y|\leq\Lambda^L_{<}$ and $|\mu_L|\leq\Lambda_\mu$ (row II), and only hole-like spinor exists when $|k_y|>\Lambda^L_{<}$ or $|\mu_L|>\Lambda_\mu$ (row I). Similarly, in the right, $n$-doped region ($\mu_R>0$), both spinors exist when $|k_y|\leq\Lambda^R_{<}$ and $|\mu_R|\leq\Lambda_\mu$ (column III), otherwise for $|k_y|>\Lambda^R_{<}$ or $|\mu_R|>\Lambda_\mu$,  only particle-like spinor exists (column IV).}
\end{table}
%%%%%%%%%%%%%%%%%%%%%%%%%%%%%%%%%%%%%%%%%%%%%%%%%%%%%%%%%%%%%%

Let us now turn to the analysis of the incoming and outgoing states. Combining all cases described in TABLE~\ref{tab:scattering type}, we can write the incoming states as 
\begin{equation}\label{Eq: incoming/outgoing states}
\begin{split}
    \vec{\Phi}^{L,+}&=\begin{pmatrix}
    \vec{\Psi}^L_p\Theta(\Lambda_{\mu}-|\mu_L|)\Theta(\Lambda^L_{<}-|k_y|)\\
    \vec{\Psi}^L_h
    \end{pmatrix}\,,\\
    \vec{\Phi}^{R,+}&=\begin{pmatrix}
    \vec{\Psi}^R_p\\
    \vec{\Psi}^R_h\Theta(\Lambda_{\mu}-|\mu_R|)\Theta(\Lambda^R_{<}-|k_y|)
    \end{pmatrix}\,,
\end{split}
\end{equation}
where the subscript $a=p/h$ in $\vec{\Psi}^l_a$ denotes the contribution of particle/hole-like spinor in the scattering state. 
The Heavy-side functions $\Theta(\Lambda^l_{<}-|k_y|)$ and $\Theta(\Lambda_{\mu}-|\mu_l|)$ take care of  the available band spinors. 

For the outgoing states $\vec{\Phi}^{l,-}$, we omit the cutoffs in order to incorporate the fact that the incoming states can also scatter into bound states with imaginary momenta that are localized at the $pn$ interface. Therefore, the outgoing state is defined as $\vec{\Phi}^{l,-}=(\vec{\Psi}^{l}_p,\vec{\Psi}^{l}_h)^T$. Unlike  the incoming states, the available band spinors $\vec{\Psi}_a^l$ are fixed by the incoming states since the $k_y$-momentum is conserved during the scattering process. The outgoing band spinors are hence obtained by searching for the intersection points between $k_y=k$ and the Fermi surfaces, where $k$ is the $k_y$-momentum of the incoming states, see Fig.~\ref{Fig:scattering angles and velocity map}(a). More formally, we obtain the $k_x$ momentum of the outgoing band spinors by solving

\begin{equation}\label{eq:kx}
    (k_x)^{2}+k^2=(\Lambda^{l}_{\lessgtr})^2 \,,
\end{equation}
where $\Lambda^{l}_{<} (\Lambda^{l}_{>})$ is chosen such that the $k_x$-momentum of spinors located at the inner (outer) branch is obtained. Note that when the solution for $k_x$ in Eq.~\eqref{eq:kx} is real, the corresponding band spinor propagates inside the system. Obtaining a finite imaginary part to $k_x$ indicates localization of the spinor at the interface.

\subsection{Scattering angles and Snell's law}
\begin{figure}[t!]
		\centering
		\includegraphics[scale=1]{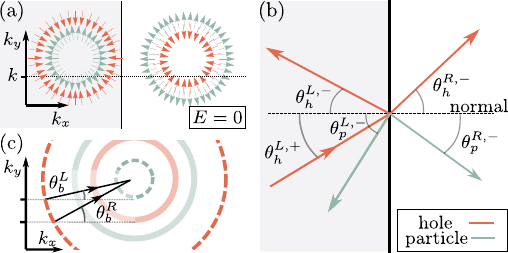}
		\caption{A sketch of the scattering angles and scattering processes. (a) The direction of the group velocities for the particle- and hole-like states at their respective Fermi surfaces, cf.~Eq.~\eqref{eq:group velocity}. With preserving $k_y$ momentum, the possible final states and their scattering angles are determined by the intersection points between the Fermi surfaces and the $k_y$ momentum of the incident state marked by the dashed line $k_y=k$, cf.~Eq.~\eqref{eq:kx}. (b) The scattering angles are defined with respect to the normal (dashed line) of the interface (solid black line), cf.~Eq.~\eqref{Eq: scattering angle}. As a hole-like state is injected from the left side at the incident angle $\theta_h^{L,+}$, the angle for the reflected particle (hole)-like state is denoted by $\theta^{L,-}_p$ ($\theta^{L,-}_h$), and the refraction angle for the transmitted particle(hole)-like states are labeled as $\theta^{R,-}_p$ ($\theta^{R,-}_h$). The particle- and hole-like states are represented by the green and red arrows aligning in the direction of their group velocities. (c) The four Fermi surfaces of the whole $pn$ junction. The dashed (solid) lines colored with green (red) represent the particle (hole)-like branch of the Fermi surface in the $p$-($n$-)doped region. The scattering processes are categorized concerning the $k_y$ momentum where the inner branch of a Fermi surface terminates. The corresponding incident angles are denoted with $\theta_b^L$ and $\theta_b^R$. We set the chemical potentials to be finite, i.e. $|\mu_L|\neq|\mu_R|\neq0$, and of opposite signs in both regions.}
		\label{Fig:scattering angles and velocity map}
\end{figure}

Let us now return to the optical-like aspects of our scattering junction and connect to standard optics. In this regard, we need to rewrite the scattering problem in terms of a propagation direction and group velocities of the wavelike states that participate in the scattering. Therefore, we make a one-to-one mapping between propagating solutions of Eq.~\eqref{eq:kx} -- given by real $k_x$ -- and their group velocity obtained as [see also Figs.~\ref{Fig:scattering angles and velocity map}(a) and (b)]
\begin{equation}\label{eq:group velocity}
    \mathbf{v} \equiv \mathbf{\nabla}_{\mathbf{k}} E(\mathbf{k})
    = \xi_a\frac{2\mathcal{M}_2\sqrt{\mu_{l}^{2}-\mathcal{A}_{c}^{2}}}{\sqrt{(\mathcal{M}_0+\mathcal{M}_2 |\mathbf{k}|^2)^2+\mathcal{A}_c^2}}\mathbf{k}\,,
\end{equation}
where $\xi_p=1$ and $\xi_h=-1$ indicate the sign of the effective mass of the spinors. In Fig.~\ref{Fig:scattering angles and velocity map}(a), we show, as a function of momentum, the group velocities of both particle- and hole-like branches in $p$- and $n$-doped regions calculated using Eq.~\eqref{eq:group velocity}.
The band spinors with $v_x>0$ in the left region are propagating towards the $pn$ interface, while the ones with $v_x>0$ in the right region propagate away from it. On the other hand, the band spinors with $v_x<0$ propagate away from the interface in the left region, and in the right region they propagate toward the $pn$ interface.

%%%%%%%%%%%%%%%%%%%%%%%%%%%%%%%%%%%%%%%%%%%%%%%%%%%%%%
%%%%%%%%%%%%%%%%%%%%%%%%%%%%%%%%%%%%%%%%%%%%%%%%%%%%%%%%%%%

As the propagating direction of the spinor $\vec{\Psi}^l_a$ is the direction of the group velocity defined in Eq.~\eqref{eq:group velocity}, for a spinor with momentum $\mathbf{k}=(k_x,k_y)$, we define the scattering angle $\theta^{l,d}_{a}$ as 
\begin{equation}\label{Eq: scattering angle}
    \theta^{l,d}_{a}=\xi_{a}\mathrm{sgn}(k_y)\tan^{-1}\left(\frac{|k_y|}{|k_x|}\right).
\end{equation}
As such, we can characterize the incident states by their incident angles to directly compare to the geometric description of traditional optics in the following discussion. In Fig.~\ref{Fig:scattering angles and velocity map}(b), we depict a scattering process where a hole-like spinor enters from the left lead at an angle $\theta^{L,+}_h$, the outgoing scattering states are marked using arrows oriented at their corresponding scattering angles $\theta^{l,-}_a$. The scattering process depicted in Fig.~\ref{Fig:scattering angles and velocity map}(b) directly follows from the equivalent momentum-space description shown in Fig.~\ref{Fig:scattering angles and velocity map}(a), i.e., the pointing directions of the velocity map in \ref{Fig:scattering angles and velocity map}(a) directly yields the scattering angles in \ref{Fig:scattering angles and velocity map}(b). 

Notice that for $|k_y|=\Lambda^{L/R}_{<}$, we can define two angles $\theta_{b}^{L/R}=\sin^{-1}(\Lambda^{L/R}_</|\mathbf{k}^{+}|)$, where $\mathbf{k}^{+}$ is the momentum of the incident states. The polarization of the propagating band spinors changes abruptly [see also Fig.~\ref{Fig:scattering angles and velocity map}(c)] at these angles, and therefore, we identify $\theta_{b}^{l}$ as an effective “Brewster's angle” of the scattering process.

Last, using the fact that the participating band spinors in Eq.~\eqref{eq:scattering equation1} are connected by the conserved $k_y$-momentum at the $pn$ junction, we can establish an effective Snell's law for the $pn$ junction once their propagating directions are known. For refractive scattering (see Appendix~\ref{sec: snell's law reflection} for the reflective case), the effective Snell's law is given by
\begin{equation}\label{Eq: Snell's law}
\begin{split}
    \Lambda_{\lessgtr}^{l}\sin(\theta^{l,+}_a)&=-\Lambda^{\bar{l}}_{\lessgtr}\sin(\theta^{\bar{l},-}_{\bar{a}})\,,\\
    \Lambda_{\lessgtr}^{l}\sin(\theta^{l,+}_a)&=\Lambda^{\bar{l}}_{\gtrless}\sin(\theta^{\bar{l},-}_{a})\,,
\end{split}
\end{equation}
where the radii of the Fermi surfaces $\Lambda^{l}_{\lessgtr}$ serve as the refractive indices. We remind the reader that the superscript $l=R/L$, from which follows that $\bar{l}=L/R$. From Eqs.~\eqref{Eq: Snell's law}, it follows that the transmission from the outer (inner) branch on the left to the outer (inner) branch on the right has a positive refractive index, while, surprisingly, the transmission processes that change the branch (outer-inner or inner-outer), involves a negative refractive index [see also Fig.~\ref{Fig:scattering angles and velocity map}(b)]. In other words, the scattering between band spinors simultaneously manifests both a normal and a negative refractive index depending on their polarizations.   
Such polarization-dependent refraction motivates us to propose an electronic bifurcating lens using the $pn$ junction in Sec.~\ref{subsec: Birfurcating lens}.

\subsection{Scattering amplitudes   \label{subsec:scattering amplitudes}}
\begin{figure}[t!]
		\centering
		\includegraphics[scale=1]{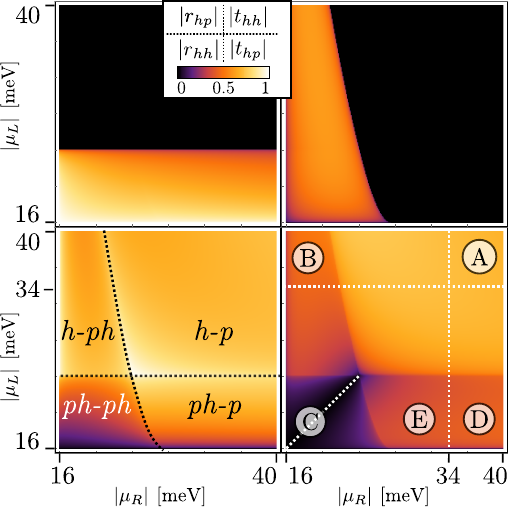}
		\caption{Scattering amplitudes for a hole-like state incident from the left side at the angle $\theta_h^{+,L}=\frac{\pi}{6}$ as a function of the chemical potentials $|\mu_L|$ and $|\mu_R|$, obtained using a scattering matrix approach. The black dashed lines mark the distinct boundary of the scattering amplitudes between different scattering regimes. In the lower right panel, the horizontal and vertical white dashed lines indicate the energy cutoff $\Lambda_\mu$ discussed in Sec.~\ref{sec:model}. The diagonal marked with a white dashed line denotes the vanishing hole-to-electron transmission $t_{hp}$ in the \textit{ph-ph} scattering regime. The encircled (A), (B), (C) and (D) mark four pairs of chemical potentials supporting scattering regime \textit{h-p}, \textit{h-ph}, \textit{ph-ph} and \textit{ph-p}, respectively. The encircled (E) marks a pair of unequal chemical potentials where both the L and R sides are in the inverted band region. In all plots, we used the following parameters: $\mathcal{M}_0=-30\ \mathrm{meV}$, $\mathcal{M}_2=700\ \mathrm{meV}$ and $\mathcal{A}_c=16\ \mathrm{meV}$.
        }	
		\label{Fig:scattering amplitues}
\end{figure}

Until now, we discussed all scattering processes between spinor states $\vec{\Phi}^{L,+}$ and $\vec{\Phi}^{l,-}$, see Eq.~\eqref{eq:scattering equation1}, without taking coefficients of the scattering matrix~\eqref{eq:s matrix} into account. Yet, they are crucial in determining the transport properties as they provide us with the probability of each process to occur. Therefore, in what follows, we study the reflection and transmission coefficients contained in matrices $\mathbf{r}$ and $\mathbf{t}$, respectively [see also Appendix~\ref{app:scattering matrix approach}]. 
Without loss of generality, we focus on a situation where a hole-like spinor enters from the left, $p$-doped, region, and solve Eq.~\eqref{eq:scattering equation1} to obtain the scattering amplitudes $|r_{hp}|$, $|r_{hh}|$, $|t_{hh}|$ and $|t_{hp}|$. 
The solutions of Eq.~\eqref{eq:scattering equation1} are shown in Fig.~\ref{Fig:scattering amplitues} as a function of chemical potentials in the $p$ and $n$ doped regions, $\mu_L$ and $\mu_R$, respectively, for an incident angle of a hole-like spinor of $\theta^{L,+}_h=\pi/6$. Note that the case where a particle-like spinor enters from the $p$-doped region is analyzed in Appendix~\ref{app:incident particle}.

Starting with the scattering processes containing states from both outer and inner branches, which give solutions for $|r_{hp}|$ and $|t_{hh}|$, we notice the appearance of distinct termination lines, see the two top panels in Fig.~\ref{Fig:scattering amplitues}. A hint is provided in TABLE~\ref{tab:scattering type} -- the scattering regime changes between case I (IV) and II (III) when the incident angle and chemical potential vary in the $p$-($n$-)doped region. This change eventually appears as the termination line in $r_{hp}\ (t_{hh})$. Here, as the incident angle is fixed, it occurs once the above-defined Brewster's angle becomes smaller than the incident angle. On a more technical level, for the hole-to-particle reflection $r_{hp}$, the termination line $\mu_L=\mu'$ is obtained using
\begin{equation}   
    \sin\left(\theta^{L,+}_h\right)=\frac{\Lambda^L_{<}}{\Lambda^L_{>}}\bigg|_{\mu_L=\mu'}\,,
\end{equation}
i.e., the chemical potential beyond which the corresponding $k_y$-momentum for the incident angle $\theta^{L,+}_h$ has no intersection point with the inner branch of the Fermi surface in the $p$-doped region.
For the incident angle $\theta^{L,+}_h=\pi/6$, the termination line is $\mu'=(9\mathcal{M}_{0}^2/25+\mathcal{A}_{c}^{2})^{\frac{1}{2}}$. In other words, the conservation condition of the $k_y$ momentum guarantees that for $|\mu_L|>\mu'$ there are no propagating particle-like states in the $p$-doped region to which the incident hole-like state can reflect.
Similarly, the termination line for $t_{hh}$ is given by $\mu''=[(4\sqrt{\mu_{R}^{2}-\mathcal{A}_c^2}+3\mathcal{M}_{0})^2+\mathcal{A}_{c}^2]^\frac{1}{2}$ for $\mu_R<\sqrt{9\mathcal{M}_{0}^2/16 +\mathcal{A}_{c}^2}$, as the hole-like states become bounded for $|\mu_L|>\mu''$. 

In contrast to the branch-changing processes described above, the scattering processes restricted to the outer branches, which yield $r_{hh}$ and $t_{hp}$, have support within the whole chosen range of the chemical potential, as shown in the lower panels of Fig.~\ref{Fig:scattering amplitues}. Such processes are more complex and have a nonmonotonous behavior as a function of the chemical potentials. 
In the lower-left panel of Fig.~\ref{Fig:scattering amplitues}, we use the black dashed lines to separate four different regimes according to the polarizations of participanting scattering states, namely \textit{ph-ph}, \textit{h-ph}, \textit{ph-h}, and \textit{h-p}. The lines of separation coincide with the $\mu'$ and $\mu''$. Note that this separation holds only for the chosen incident angle $\theta^{L,+}_h=\pi/6$. If one takes all possible incident angles into account, the separation between the different regimes is given by the white horizontal and vertical dashed lines, see the lower-right panel of Fig.~\ref{Fig:scattering amplitues}, which mark the energy cutoff $\Lambda_{\mu}$. In this case, the regimes \textit{h-p}, \textit{h-ph}, \textit{ph-ph} and \textit{ph-p} are marked with encircled (A), (B), (E) and (D), respectively, while (C) is lying in the (E) region and marks the line where $|\mu_L|=|\mu_R|$.

Let us return to the case of a single incident hole with $\theta^{L,+}_h=\pi/6$. Generally, for the potentials close to the main gap, the scattering is dominated by reflections, mostly by $r_{hp}$. For simultaneously high chemical potentials, i.e., when $|\mu_L|$ and $|\mu_R|$ are in the \textit{h-p} regime, the scattering processes are entirely determined by the amplitudes $|r_{hh}|$ and $|t_{hp}|$. The branch-changing transmission amplitude $|t_{hp}|$ shows an interesting structure in $\mu_L - \mu_R$ parameter space, namely while being maximized in the \textit{h-p} region, it is heavily suppressed in the \textit{ph-ph} regime while it completely vanishes when $|\mu_L|=|\mu_R|$, see (C) in Fig.~\ref{Fig:scattering amplitues}.

To explain such behavior of $|t_{hp}|$, we first concentrate on a symmetric $pn$ junction with $\mu_L=-\mu_R$ (for the asymmetric case, see Appendix~\ref{sec:app_asmmetric_junction}), which hosts only \textit{ph-ph} and \textit{h-p} regimes. For the former regime, all $k_x$ momenta involved in scattering are real, and the scattering states propagate through the system. In this situation, we can analytically obtain the scattering amplitudes:
\begin{align}\label{eq: amplitudes symm}
         r_{hh}&=\frac{k_x^>-k_x^<}{k_x^>+k_x^<}, &
         t_{hp}&=0,
         \\
         r_{hp}&=-\frac{2\sqrt{k_x^>k_x^<}}{k_x^>+k_x^<}\frac{\mathcal{A}_c}{\mu}\,,
         &
         t_{hh}&=\frac{2\sqrt{k_x^>k_x^<}}{k_x^>+k_x^<}\sqrt{1-\frac{\mathcal{A}_c^2}{\mu^2}}\,,\nonumber
 \end{align}
where $k_x^{\lessgtr}=[(\Lambda^{l}_{\lessgtr})^2-k_{y}^{2}]^{\frac{1}{2}}$ and $|\mu_L|=|\mu_R|=\mu$. We notice no polarization-flipping transmission in the \textit{ph-ph} regime, i.e., $t_{hp}=0$, which explains the effect marked with (C) in Fig.~\ref{Fig:scattering amplitues}.

On the other hand, in the \textit{h-p} regime of the symmetrically-doped $pn$ junction, the inner branch is missing on both sides of the $pn$ interface, leading to bound states with imaginary $k_x$ momenta. The presence of such bound states modifies the scattering processes between the propagating states by introducing a correction term to the scattering amplitudes, as shown in Appendix~\ref{sec:app_bound_states}. For example, the hole-to-particle transmission $t_{hp}^{BS}$ due to the bound states reads
 \begin{equation}
     t_{hp}^{BS}=\frac{4k_x^>k_x^<}{(k_x^>)^2-(k_x^<)^2}\sqrt{1-\frac{\mathcal{A}_c^2}{\mu^2}}\frac{\mathcal{A}_c}{\mu}\,,
 \end{equation}
where $k^{<}_x=[-(\Lambda^{l}_{\lessgtr})^2+k_{y}^2]^{\frac{1}{2}}$ is the imaginary part of the $k_x$-momentum of the bound states. Similar corrections coming from bound states alter all scattering amplitudes in the \textit{h-p}, \textit{h-ph} and \textit{ph-p} regimes, as we show in Appendix~\ref{sec:app_bound_states}. 
\begin{figure}[ht!]
		\centering
		\includegraphics[scale=1]{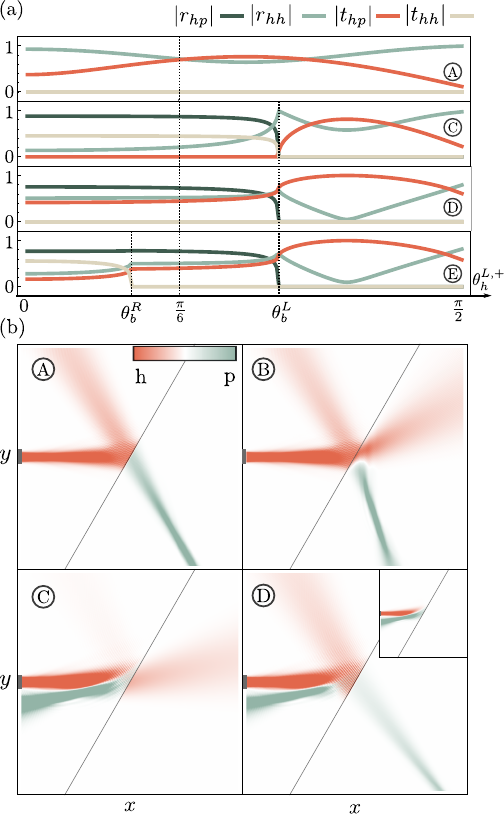}
		\caption{(a) Scattering amplitudes as a function of the incident angle $\theta_h^{L,+}$ for the chemical potentials marked in Fig.~\ref{Fig:scattering amplitues}. The scattering amplitudes show discontinuity at the Brewster angles denoted by $\theta_b^L$ and $\theta_b^R$. 
        (b) Local densities of the polarization (LDOP) from Eq.~\eqref{eq:LDOP}, i.e., the local density of band spinors, for hole-like modes injected from a lead attached in the middle of the left side of the sample, obtained numerically by solving Eq.~\eqref{eq:scattering equation1} in real space using Kwant~\cite{Groth_2014}. The interface is tilted so that the incident angle $\theta_h^{L,+}=\pi/6$. The parameters are chosen the same as in Fig.~\ref{Fig:scattering amplitues} and the corresponding chemical potentials for plots marked with (A), (B), (C), (D) and (E) are $(\mu_L,\mu_R)\in\{(-38,38),(-38,18),(-18,18),(-18,38), (-18,30)\}$ in meV-s, respectively. In the inset of (D), we used $(\mu_L,\mu_R) = (-18,0)$.
        In all numerical calculations, we used large systems with $L_x=L_y$ sites with a narrow lead (marked with a gray rectangle) of width $L_y/15$ sites attached to the left side of the system.}
		\label{Fig:scattering amplitudes for angles and local densities}
\end{figure}

Let us now expand the analysis to arbitrary incident angles. We show in Fig.~\ref{Fig:scattering amplitudes for angles and local densities}(a) the scattering amplitudes as a function of the incident angle for selected combinations of the chemical potentials marked in Fig.~\ref{Fig:scattering amplitues}. 
When the Fermi surface in both regions lies outside the inverted-band regime, e.g., as in case (A), no Brewster's angle exists and all the scattering amplitudes are smooth functions of the incident angle. 
Reducing the $|\mu_L|$ below $\Lambda_{\mu}$, Brewster's angles appear and introduce the discontinuity in the scattering amplitudes, as is seen at (C) and (D) in Fig.~\ref{Fig:scattering amplitudes for angles and local densities}(a) -- both of which show a single discontinuity point. In the symmetric case marked by (C), the two Brewster's angles are equal $\theta_b^L=\theta_b^R$, and therefore, the discontinuity in the scattering amplitudes occurs only once. Similarly, a single discontinuity appears in (D), but this time due to the presence of only one Brewster's angle -- a consequence of the Fermi surface in the $n$-doped region containing only one branch.
The most complex behavior can be seen in (E), where $|\mu_L| \neq |\mu_R|$ and $|\mu_L|,\ |\mu_R|<\Lambda_\mu$ [see also Fig.~\ref{Fig:scattering amplitues}], where both Brewster's angles $\theta_b^L$ and $\theta_b^R$ exist and are non-degenerate. In that case, for the incident angle $\theta^{L,+}_{h}<\theta_b^R$, the scattering process is in the \textit{ph-ph} regime. When $\theta^{L,+}_h$ increases above $\theta_b^R$, $t_{hh}$ vanishes since the hole-like branch in the $n$-doped region does not intersect with the corresponding incident momentum $k_y$, and the scattering type is now \textit{ph-p}. Increasing the incident angle beyond $\theta_b^L$ leads to a vanishing particle branch in the $p$-doped region, and therefore, $|r_{hp}|$ drops to zero.

One more remarkable feature observed in Fig.~\ref{Fig:scattering amplitudes for angles and local densities}(a) is the evidence of  a Klein-tunnelling alike effect: close to perfect transmission occurs while all reflection coefficients become almost zero. Unlike graphene, where the Klein tunnelling occurs only for the zero-incident angle, in the inverted-band $pn$ junctions the transmissions acquiring a negative refractive index, e.g., $t_{hp}$, maximize at a finite angle, see (D) and (E) case of Fig.~\ref{Fig:scattering amplitudes for angles and local densities}, where $|t_{hp}|$ is maximized at the incident angle $\theta_b^L < \theta_h^{L,+} < \pi/2$. In the next section of this paper, we discuss in detail how this feature facilitates interesting potential experimental applications.

%%%%%%%%%%%%%%%%%%%%%%%%%%%%%%%%%%%%%%%%%%%%%%%%%%%%%%%%%%%%%
%%%%%%%%%%%%%%%%%%%%%%%%%%%%%%%%%%%%%%%%%%%%%%%%%%%%%%%%%%%%%
\section{Numerical analysis  \label{sec:numerical analysis}}
In order to verify our analytical predictions above, we develop and harness a numerical toolbox based on the Kwant software~\cite{Groth_2014}. We use a two-dimensional square lattice containing $L_x=L_y$ sites in both $x$- and $y$-direction. The unit cell size is chosen sufficiently small such that the low-energy physics is effectively described as in the continuum system. By adding positive (negative) onsite potentials onto specific regions in the numerical “sample”, we can emulate the effect of gating to obtain the $pn$ junction discussed in the previous sections. In other words, we can study the transport for different choices of the chemical potentials $\mu_l$. In order to study the transport, we attach a narrow semi-infinite source lead in the x-direction to the middle of the left side of the system. The lead has a finite width of $L_y/15$ sites in the y-direction, which serves as an electron source, see Fig.~\ref{Fig:scattering amplitudes for angles and local densities}(b). As drain,  we attach semi-infinite leads around the whole system such that all spinors leave the system without additional reflections at the boundaries. Finally, all leads are described by the same Hamiltonian as their adherent regions.

Trajectories of injected states, as well as the ones after the scattering, are captured by the local density of the polarization (LDOP), which is defined as 
\begin{align}   \label{eq:LDOP}
    {\rm LDOP}(\mathbf{x})=\sum_{n=1}^N \vec{\Psi}_n^\dagger(\mathbf{x})\sigma_z\vec{\Psi}_n(\mathbf{x})\,,
\end{align}
where $\vec{\Psi}_n(\mathbf{x})$ is the wave function of the $n$-th channel injected from the attached lead, and the sum runs over all discrete channels in the lead that form in the y-direction. Note that the effect of those discrete channels is the same as having discrete $k_y$ momenta in discussion regarding previous sections.
The numerically calculated LDOP is shown in Fig.~\ref{Fig:scattering amplitudes for angles and local densities}(b) for the case where hole-like states are injected from the left lead. We chose several different combinations of the chemical potentials from the regions marked in Fig.~\ref{Fig:scattering amplitues}. 

%%%%%%%%%%%%%%%%%%%%%%%%%%%%%%%%%%%%%%%%%%%%%%%	
\subsection{Negative refraction}
As discussed in Eqs.~\eqref{Eq: Snell's law}, negative diffraction can occur in branch-changing scattering processes. Therefore, there are two types of negative diffraction in our system; one in the transmission process and the other in the reflection process. For the case of an incident hole-like state, they manifest in $t_{hp}$ and $r_{hp}$, respectively. We observe both types of negative refraction in the numerical simulation, see two left panels of Fig.~\ref{Fig:scattering amplitudes for angles and local densities}(b) marked with (A) and (C). The former is in the \textit{h-p} regime, where we observe negative refraction in transmission. 
The latter is in the \textit{ph-ph} regime, and we observe almost perfect reflection of the hole-like states to particle-like ones with negative refraction in reflected scattering, see also Appendix~\ref{sec: snell's law reflection}.

%%%%%%%%%%%%%%%%%%%%%%%%%%%%%%%%%%%%%%%%%%%%%%%%%%%%%%%%%%%%%
%%%%%%%%%%%%%%%%%%%%%%%%%%%%%%%%%%%%%%%%%%%%%%%%%%%%%%%%%%%%%
\section{Electron optics devices}
\label{sec:devices}
Using the scattering theory developed in the previous section and the knowledge that comes from it, we can now numerically study possible applications of the rich scattering processes in the inverted-band $pn$ junctions. 
In a recent work~\cite{Karalic2020}, some of us showed that the inverted-band $pn$ junctions provide a novel mechanism for realizing electron optics. The electron-hole interference pattern is observed in a Fabry-P\'{e}rot interferometer based on an inverted-band $pnp$ junction. Here, we expand on the variety of potential applications in terms of geometrical optics, where our understanding of the angular dependence of the scattering process paves the way to design fundamental optical components for electronic systems, such as mirrors and focusing lens. Moreover, since the band spinor in the inverted-band systems plays the same role as the polarization of photons in traditional optical systems, by analogy to the optical components, we can construct bifurcating lens and polarizers for the band spinors using different $pn$ junctions.

%%%%%%%%%%%%%%%%%%%%%%%%%%%%%%%%%%%%%%%%%%%%%%%	
\subsection{Bifurcating lens    \label{subsec: Birfurcating lens}}
We start by discussing the effects that occur in non-symmetric $pn$ junctions, i.e., when $|\mu_L| \neq |\mu_R|$. 
The first effect is bifurcation. In optical birefringent materials, the refraction index differs for photons with different polarizations and it is possible to spatially separate photons with different polarizations.
Equivalently, in our electronic system where the alignment of the band spinors plays the role of polarization, we can obtain different refraction indices for particle- and hole-like spinors, realizing with that an electronic bifurcating lens for the band polarization. One such realization is shown in the upper right panel of Fig.~\ref{Fig:scattering amplitudes for angles and local densities}(b) marked with (B): a horizontal propagating hole-like state hits a right-tilted interface in the \textit{h-ph} regime. The refracted particle-like states (colored green) propagate to the lower-right direction, while the hole-like states (colored red) head to the upper-right corner. Therefore, the inverted-band $pn$ junction automatically realizes a bifurcating lens for the band spinors, due to the fact that the polarization-preserving transmission acquires a positive refraction index while the polarization-flipping transmission acquires a negative refraction index.

%%%%%%%%%%%%%%%%%%%%%%%%%%%%%%%%%%%%%%%%%%%%%%%		
\subsection{Polarizer}
By tuning the chemical potential in the right region beyond the inverted-band regime, i.e., $|\mu_R|>\Lambda_\mu$, only the outer branch of the Fermi surface is accessible during the scattering process, meaning that only particle-like states can scatter through the $n$-doped region. In the lower right panel of Fig.~\ref{Fig:scattering amplitudes for angles and local densities}(b), marked with (D), we observe a particle-like polarizer based on such a setting. 
On the other hand, a hole-like polarizer can be built by setting $\mu_R<-\Lambda_\mu$ so that only hole-like states are transmitted to the right through the interface despite the polarization of the incident states from the left side.

%%%%%%%%%%%%%%%%%%%%%%%%%%%%%%%%%%%%%%%%%%%%%%%	
\subsection{Mirror   \label{subsec: Mirror}}
The most fundamental component in optics, i.e., the mirror is also realiseable in our setup. By tuning the chemical potential $\mu_R$ to reside inside the gap, the states entering from the left, $p$-doped region cannot scatter into the right region due to the absence of propagating states. Therefore, all incident states are reflected.
We demonstrate this in the inset of case (D) in Fig.~\ref{Fig:scattering amplitudes for angles and local densities}(b). The incident hole-like states from the left region are reflected into the particle-like states with a negative reflection angle, due to the negative refraction. Furthermore, we find by solving the scattering amplitudes that as long as the incident angle is smaller than Brewster's angle, i.e., when both the particle- and hole-like states are propagating in the $p$-doped region, the polarization-flipping reflection $r_{hp}$ dominates [see also Fig.~\ref{Fig:band spinor and bound states}(c)]. Thus, by manipulating the incident angle, we can control the polarization of the reflected states as well as their reflection angle using the inverted-band $pn$ junction mirror.

%%%%%%%%%%%%%%%%%%%%%%%%%%%%%%%%%%%%%%%%%%%%%%%%%%%%%%%%%%%%%
%%%%%%%%%%%%%%%%%%%%%%%%%%%%%%%%%%%%%%%%%%%%%%%%%%%%%%%%%%%%%
\subsection{Veselago and focusing lens    }
We turn now to discuss two additional electron-optics devices that can be constructed based on the negative refraction discussed in the previous sections. For simplicity, we restrict our discussion to a symmetric junction tuned to the \textit{h-p} regime. In this case, only polarization-flipping transmission $t_{hp}$ is allowed, and the incident angle equals the refraction angle. We use the same numerical methodology but with geometries adapted to observe the lensing effects. 

\begin{figure}
		\centering
		\includegraphics[scale=1]{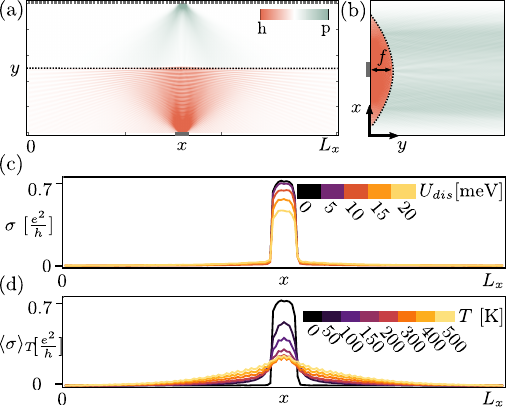}
		\caption{(a) LDOP [cf.~Eq.~\eqref{eq:LDOP}] in a Veselago lens setup based on a symmetric $pn$ junction. The hole-like states (red) are injected from the lead attached to the lower side of the sample. After the \textit{h-p} scattering at the interface (dashed line), the states transmit as particle-like (green) and focus on a point at the upper side. For the numeric calculation, we use the same parameters as in Fig.~\ref{Fig:scattering amplitues}, with which the system length $L_x$ is $2000\ \text{nm}$. (b) LDOP in a focusing lens based on a symmetric $pn$ junction. The curved interface (dashed line) transfers states injected from a point source on the left to parallel propagating beams on the right. The focusing length $f$ indicates the distance between the source and the interface. In (c) and (d), the conductance $\sigma$ through the Veselago lens is shown as a function of the position $x$ and for different strengths of disorder and at different temperatures, respectively. 
		}
		\label{Fig:devices focusing lens}
\end{figure}

In Fig.~\ref{Fig:devices focusing lens}(a), we show a Veselago lens~\cite{veselago1968,pendry2000} realized using the inverted-band $pn$ junction. We numerically calculate the LDOP~\eqref{eq:LDOP} for the case where the hole-like states are injected from a narrow source lead attached at the lower side of a flat interface. Since the $pn$ junction is symmetric, we observe a strong focusing effect of particle-like states on the other side of the interface. The focal point is located symmetrically with respect to the interface. 

We proceed now with a curved interface, with which we can realize a focusing lens, see Fig.~\ref{Fig:devices focusing lens}(b). Injecting the hole-like states from a point-like source located in the focal point on the left side of the curved interface, parallel beams of particle-like states are obtained on the right side -- without any additional spatial confinement. Equivalently, one can inject particle-like states that propagate in parallel from the right and focus them as hole-like states on the left side of the interface. To obtain the above effect, for the symmetric $pn$ junction discussed in Fig.~\ref{Fig:devices focusing lens}(b), the interface needs to have a parabolic shape defined by $x=\pm y^2/(4f)$~\cite{liu_creating_2017}, where $f$ is the focusing length indicating the distance between the lead and the interface.

%%%%%%%%%%%%%%%%%%%%%%%%%%%%%%%%%%%%%%%%%%%%%%%	
%%%%%%%%%%%%%%%%%%%%%%%%%%%%%%%%%%%%%%%%%%%%%%%%%%%%%%

%%%%%%%%%%%%%%%%%%%%%%%%%%%%%%%%%%%%%%%%%%%%%%%%%%%%%%%%%%%%%
%%%%%%%%%%%%%%%%%%%%%%%%%%%%%%%%%%%%%%%%%%%%%%%%%%%%%%%%%%%%%
\section{Effects of disorder and finite temperature}
\label{sec:temperature and disorder}
So far, our analysis was conducted in a clean system at zero temperature. In reality, however, random impurities in the junction and thermal fluctuations can be detrimental to the optical properties of the devices. In the following, we take the Veselago lens shown in Fig.~\ref{Fig:devices focusing lens}(a) as an example and verify its robustness against disorder and thermal fluctuations using our numerical simulations.

%%%%%%%%%%%%%%%%%%%%%%%%%%%%%%%%%%%%%%%%%%%%%%%%%%%%%%%
\subsection{Robustness against disorder}
First, we consider random impurities in both $p$- and $n$-doped regions and study the performance of the Veselago lens as a function of their strengths. We model realistic impurities by introducing a randomly distributed onsite potential $\mu_{dis}$, which obeys a normal distribution with a zero mean and the standard deviation $U_{dis}$. The strength of the disorder is then controlled by $U_{dis}$. Next, we fix the lower lead with a length of $L_b\approx L_x/18$ to the middle of the $p$-doped region. This lead acts as a source for the hole-like states. Furthermore, we attach a series of smaller leads each with length $L_t\approx L_x/161$ on the top of the $n$-doped region, which serves as a drain for particle-like states. 

In such a setup, we calculate the non-local conductance as the function of the position of the upper lead, which is defined as 
\begin{equation} \label{eq: condusctance}
    \sigma = \frac{e^2}{h}\sum_{n=1}^{N}\sum_{m=1}^{M}\Tr\big[t_{nm}^\dagger t_{nm}\big]\,,
\end{equation}
where $t'_{nm}$ is the modified transmission matrix defined in Eq.~\eqref{eq:s matrix} that connects the channel $n$ of the lower lead to the channel $m$ of the upper lead. We remind the reader that the discretization of $n$ and $m$ arises from the finite size of the leads. Therefore, to calculate the total transmission of injected states on the bottom to the ones that enter the top leads, one needs to sum over the contributions of all the states injected from source S and those arriving in drain D.
In Fig.~\ref{Fig:devices focusing lens}(c), we show the conductance as a function of the horizontal position of the upper lead and for different disorder amplitudes. When the system is disorder-free, the conductance is maximal when the two leads are symmetrically positioned with respect to the $pn$ interface and vanishes when the upper lead is positioned far away from the middle. Increased disorder strength reduces $\sigma$, but the maxima in the middle remain visible even at strong disorder $U_{dis}/|\mathcal{M}_0|=2/3$.
% Note that instead of maximizing at the middle, the conductance has maxima when the upper lead is slightly dislocated from the symmetric position. This is because of the finite width of the leads and the peculiar angular dependence of $t_{hp}$, which maximizes at finite angles, see the (A) case in Fig.~\ref{Fig:scattering amplitudes for angles and local densities}(a).

%%%%%%%%%%%%%%%%%%%%%%%%%%%%%%%%%%%%%%%%%%%%%%%%%%%%%%%%%%%%%
\subsection{Robustness against finite temperature}
Next, we calculate the conductance $\sigma$ at finite temperatures for the same setup using
\begin{equation}
    \langle\sigma\rangle_T = \frac{1}{2k_B T}\int^{\infty}_{-\infty}\mathrm{d}\varepsilon\frac{\sigma(\varepsilon)}{\cosh(\varepsilon/k_BT)+1}\,,
\end{equation}
where $k_B$ is the Boltzmann constant and $\sigma(\varepsilon)$ is the non-local conductance obtained using Eq.~\eqref{eq: condusctance} with a Fermi surface at $\varepsilon$. We approximate the integral using a Riemann sum, using a summation over $\sigma(\varepsilon)$ with $\varepsilon$ sampled 50 times obeying the probability distribution $P(\varepsilon)=[2k_BT(\cosh(\varepsilon/k_BT)+1)]^{-1}$.
The results are shown in Fig.~\ref{Fig:devices focusing lens}(d), where we observe that although the maximal value of the conductance drops rapidly as the temperature increases, the uneven spatial distribution of $\sigma$, i.e. the presence of the peak, is not destroyed even at high temperatures of $500$K. Unlike the disordered case, the thermal fluctuations “blur out” the focusing point while simultaneously increasing the conductance at the surrounding top leads.

%%%%%%%%%%%%%%%%%%%%%%%%%%%%%%%%%%%%%%%%%%%%%%%%%%%%%%%%%%%%%
%%%%%%%%%%%%%%%%%%%%%%%%%%%%%%%%%%%%%%%%%%%%%%%%%%%%%%%%%%%%%
\section{Conclusion and Outlook}
In this work, we developed a scattering theory for band-inverted $pn$ junctions, in which the band hybridization and  inversion yield a sombrero-hat dispersion for the band spinors, see Fig.~\ref{Fig:setup and band}. 
The sombrero-hat dispersion entails the presence of particle- and hole-like states in the system, which then correspond to the polarization of light in the language of standard optics -- with an additional difference being a nonlinear dispersion of electronic states, see Fig.~\ref{Fig:scattering processes}. In section~\ref{sec: snell's law reflection}, we showed how the scattering problem given by Eq.~\eqref{eq:scattering equation1} can be mapped to an effective Snell's law for band spinors, see also Eqs.~\eqref{Eq: Snell's law}. This allowed us to introduce -- for the first time in electronic systems -- the concept of electronic Brewster's angles considering the alignment of the band spinors as the polarization of the quasiparticles, see Fig.~\ref{Fig:scattering angles and velocity map}(b). 

Moreover, by studying Eqs.~\eqref{Eq: Snell's law}, we postulated and later observed negative refraction for branch-changing scattering processes, both in the case of transmitted and the reflected states, see Figs.~\ref{Fig:scattering amplitudes for angles and local densities}(b) and~\ref{Fig:devices focusing lens}(a).
By further exploring the parameter space, i.e., by tuning the chemical potentials in the $p$-doped and $n$-doped region, we find combinations based on which the fundamental optical components are realizable, such as mirrors, bifurcating lens, polarizers, and focusing lens. We verify the feasibility of realizing these devices using numerical simulations, see Fig.~\ref{Fig:scattering amplitudes for angles and local densities}(b). Using such optical components, we show in Figs.~\ref{Fig:devices focusing lens}(a) and (b) realizations of Veselago and focusing lenses. Additionally, for the former, we discuss the influence of disorder and thermal fluctuations -- the two main sources of decoherence in realistic systems -- on the performance of the devices, see Figs.~\ref{Fig:devices focusing lens}(c) and (d).
The robustness shown in those results supports experimental feasibility and confirms that the inverted-band $pn$ junction is a promising platform for realizing electron optics phenomena.

We emphasize that the direct observation of one of the effects we discuss in this manuscript -- the negative refraction -- in mesoscopic 2D systems was accomplished recently using specially engineered polaritonic platforms~\cite{Sternbach2023,Hu2023}. The only known solid state platform, where some aspects of 2D electron optics have been proposed~\cite{Cheianov2007} and realized~\cite{Chen2016, rickhaus2015}, are mostly the graphene-based devices. 
However, to observe more complicated optical effects like Veselago lensing in graphene-based $pn$ junctions~\cite{Cheianov2007}, the interface between $p$- and $n$-doped regions has to be perfectly smooth in $y$-direction and extremely narrow. Otherwise, for the graded $pn$ interface, the transmission at finite angles is exponentially suppressed and only electrons impinging the interface at zero angle transmit thanks to the Klein-tunneling effect~\cite{cheianov_selective_2006}. 
As a result, the graphene-based $pn$ junction acts as a bad collimator with a drastically reduced beam intensity. This makes direct observation of novel electron-optics properties, such as negative refraction, with standard graphene devices experimentally unachievable.
In such a case, the focusing effects can only be indirectly observed in experiments using a magnetic field to bend the trajectory of the electrons~\cite{Chen2016}. 

In inverted-band $pn$ junctions studied in this work, crucially, the transmission amplitude that generates the negative refractive index maximizes at a finite angle, and therefore, the effects and devices discussed in our manuscript could survive even in the presence of a graded interface. 
In other words, the Klein tunneling alike effect that we report in Fig.~\ref{Fig:scattering amplitudes for angles and local densities}(a) enables direct experimental observation of the negative refractive index, and consequently, the realization of the devices discussed in sections~\ref{sec:numerical analysis} and~\ref{sec:devices}.
Thus, we expect our proposal to be directly measured in real materials such as InAs/GaSb, see Ref.~\cite{Karalic2020} where an electronic Fabry-P\'{e}rot interferometer has been theoretically proposed and experimentally investigated, as well as in bilayer graphene and other topological materials exhibiting band inversion.

We note that our analysis is quite general and did not aim to include all microscopic details of the dispersion of specific materials. Based on our description, it is straightforward to generalize our multi-band scattering theory to other systems featuring more complicated inverted-band structures, e.g., to systems with different effective masses for particle- and hole-like states, where different group velocities of particle- and hole-like states can result in rich interference patterns. In addition, our work can be applied to topologically non-trivial systems, where the topological phase transition can thoroughly change the angular dependence of the scattering amplitudes.
Since the negative refraction in our model relies on the sombrero-shaped band structure of the $pn$ junctions, it can also manifest in a spin-full system, where the Hamiltonian is modified by spin-orbit coupling: as long as the required band structure is preserved for at least one spin direction, our predictions will hold.
Lastly, our result facilitates the design of, other, more complicated electron-optics devices, such as interferometers or electron spectrometers, which can be constructed using inverted-band $pn$ junctions using the basic components we propose.

%%%%%%%%%%%%%%%%%%%%%%%%%%%%%%%%%%%%%%%%%%%%%%%%%%%
\begin{acknowledgments}
This work was supported by the Swiss National Science Foundation (A.\v{S}. Grant No.~199969), the Deutsche Forschungsgemeinschaft (DFG) through project number 449653034, and ETH research grant ETH-28 23-1. Our numerical simulations are performed using the python package Kwant~\cite{Groth_2014}. 
\end{acknowledgments}
%%%%%%%%%%%%%%%%%%%%%%%%%%%%%%%%%%%%%%%%%%%%%%%%%%%%%%
%%%%%%%%%%%%%%%%%%%%%%%%%%%%%%%%%%%%%%%%%%%%%%%%%%%%%%
\appendix
%%%%%%%%%%%%%%%%%%%%%%%%%%%%%%%%%%%%%%%%%%%%%%%%%%%%%%
%%%%%%%%%%%%%%%%%%%%%%%%%%%%%%%%%%%%%%%%%%%%%%%%%%%%%%
\section{Details on the scattering matrix approach \label{app:scattering matrix approach}}
To obtain the corresponding amplitudes for transmissions and reflections at the interface, we employ a scattering matrix approach in Sec.~\ref{subsec:scattering amplitudes}. As defined in Eq.~\eqref{eq:scattering equation1}, the scattering matrix in our system consists of four scattering amplitudes being $2\times2$ matrices
\begin{align}\label{eq:scattering amplitudes}
        \mathbf{r}&=\begin{pmatrix}
        r_{pp} & r_{ph}\\
        r_{hp} & r_{hh}
        \end{pmatrix}, &
        \mathbf{t}&=\begin{pmatrix}
        t_{pp} & t_{ph}\\
        t_{hp} & t_{hh}
        \end{pmatrix},\nonumber\\
        \mathbf{r}'&=\begin{pmatrix}
        r'_{pp} & r'_{ph}\\
        r'_{hp} & r'_{hh}
        \end{pmatrix}, &
        \mathbf{t}'&=\begin{pmatrix}
        t'_{pp} & t'_{ph}\\
        t'_{hp} & t'_{hh}
        \end{pmatrix},
\end{align}
where each entry in the matrices indicates the probability amplitude for the corresponding scattering process to occur. For given parameters, their values are determined using the continuity condition of the wavefunctions at the interface, i.e., the wavefunctions for $p$- and $n$-doped region and their first derivative must be the same value at the interface. For instance, to obtain $t_{ha}$ and $r_{ha}$, where $a=h,p$, we assume a hole-like spinor with $k_y=k$ impinging the interface aligned in the $y$-direction from the left. The wavefunctions in the $p$ and $n$ regions are thus given by
{\footnotesize
\begin{equation}
\begin{split}
    \vec{\Psi}^{L}(\mathbf{x},k)&=\vec{\Psi}^{L}_{h}e^{-i\mathbf{k}_{h}^{L,+}\mathbf{x}}+\tilde{r}_{hp}\vec{\Psi}^{L}_{p}e^{-i\mathbf{k}_{p}^{L,-}\mathbf{x}}+\tilde{r}_{hh}\vec{\Psi}^{L}_{h}e^{-i\mathbf{k}_{h}^{L,-}\mathbf{x}},\\
    \vec{\Psi}^{R}(\mathbf{x},k)&=\tilde{t}_{hp}\vec{\Psi}^{R}_{p}e^{-i\mathbf{k}_{p}^{R,-}\mathbf{x}}+\tilde{t}_{hh}\vec{\Psi}^{R}_{h}e^{-i\mathbf{k}_{h}^{R,-}\mathbf{x}},
\end{split}
\end{equation}}
where $\mathbf{k}^{l,d}_{a}=(k^{l,d}_{a},k)^\mathrm{T}$ is the momentum of spinors with $k_y=k$ and $k_x=k^{l,d}_{a}$ obtained by solving Eq.~\eqref{eq:kx}. Next, as the spinor $\vec{\Psi}^{l,d}_{a}$ is a two-component vector, we have four equations by implementing the continuity condition at the $pn$ interface, i.e., at $x=0$
\begin{equation}\label{eq: continuity condition}
    \begin{split}
        \vec{\Psi}^{L}(\mathbf{x},k)\big|_{x=0}&= \vec{\Psi}^{R}(\mathbf{x},k)\big|_{x=0},\\
        \partial_x\vec{\Psi}^{L}(\mathbf{x},k)\big|_{x=0}&= \partial_x\vec{\Psi}^{R}(\mathbf{x},k)\big|_{x=0},
    \end{split}
\end{equation}
with which the amplitudes of $\tilde{t}_{ha}$ and $\tilde{r}_{ha}$ can be determined. Yet, they are not the scattering amplitudes in Eq.~\eqref{eq:scattering amplitudes} as the conservation of the spinor current in the direction of the interface normal is not guaranteed by the continuity conditions. Therefore, we further modify the amplitudes $\tilde{t}_{ha}$ and $\tilde{r}_{ha}$ with the ratio of group velocities of the initial and final states in the direction of the interface normal, i.e., $x$-direction. As an example, $t_{hp}$ is defined as 
\begin{equation}
t_{hp}=\sqrt{\frac{v_p^R}{v_h^L}}\tilde{t}_{hp},
\end{equation}
where $v_a^l$ is the $x$-component of the group velocity defined in Eq.~\eqref{eq:group velocity} for the corresponding spinor $\vec{\Psi}^l_a$. Moreover, when bound states are involved, $[v_a^l]_x=0$ as they are localized at the interface. Other components in Eq.~\eqref{eq:scattering amplitudes} are obtained by repeating this procedure for the incident particle-like spinor, as well as using the reciprocal relation between different scattering processes. 

%%%%%%%%%%%%%%%%%%%%%%%%%%%%%%%%%%%%%%%%%%%%%%%%%%%%%%
%%%%%%%%%%%%%%%%%%%%%%%%%%%%%%%%%%%%%%%%%%%%%%%%%%%%%%
\section{Effective Snell's law for the reflective scattering}\label{sec: snell's law reflection}
As the $k_y$-momentum is preserved during the scattering process, we obtain an effective Snell's law for reflected states as
\begin{align}
    \Lambda_{\lessgtr}^{l}\sin(\theta^{l,+}_a)&=\Lambda^{l}_{\lessgtr}\sin(\theta^{l,-}_{a}),\\
    \Lambda_{\lessgtr}^{l}\sin(\theta^{l,+}_a)&=-\Lambda^{l}_{\gtrless}\sin(\theta^{l,-}_{\bar{a}}),
\end{align}
where the minus sign for polarization-flipping reflection $r_{hp}$ and $r_{ph}$ indicates that the reflected states that acquire opposite polarization propagate in the same sector -- if one imagines the $x$- and $y$-axis to divide the space into four sectors -- as the incident states. Moreover, for the symmetric $pn$ junction, we obtain $\theta^{l,+}_{a}=-\theta^{l,-}_{\bar{a}}$, which suggests that the reflected states propagate backwards along the trajectory of the incident states, see also in Fig.~\ref{Fig:setup and band} and case (C) in Fig.~\ref{Fig:scattering amplitudes for angles and local densities}.

%%%%%%%%%%%%%%%%%%%%%%%%%%%%%%%%%%%%%%%%%%%%%%%%%%%%%%
%%%%%%%%%%%%%%%%%%%%%%%%%%%%%%%%%%%%%%%%%%%%%%%%%%%%%%
\section{Incident particle-like spinor  \label{app:incident particle}}
%%%%%%%%%%%%%%%%%%%%%%%%%%%%%%%%%%%%%%%%%%%%%
\begin{figure}
		\centering
		\includegraphics[scale=1]{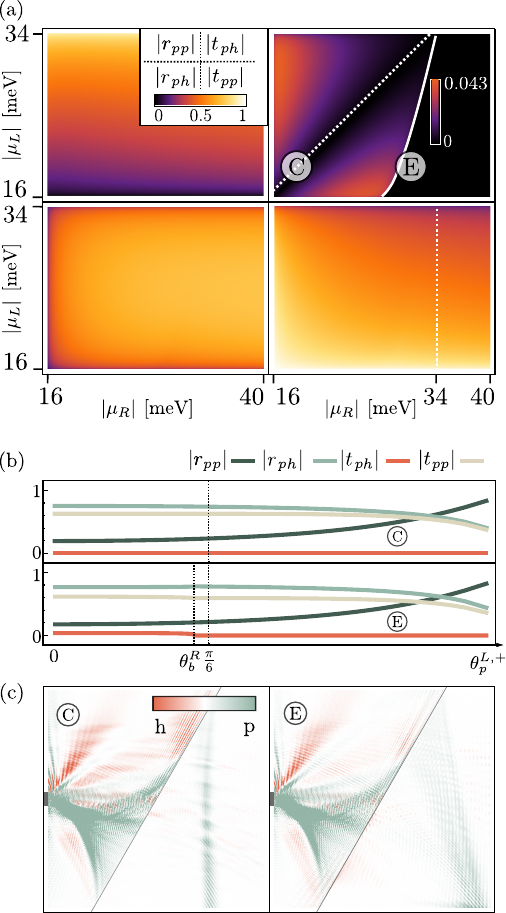}
		\caption{(a) Scattering amplitudes for a particle-like state incident from the left side at the angle $\theta_h^{+,L}=\frac{\pi}{6}$ as a function of the chemical potentials $|\mu_L|$ and $|\mu_R|$, obtained using a scattering matrix approach. The solid white line in the upper right panel indicates the boundary between the scattering regime \textit{ph-ph} (to the left) and \textit{ph-p} (to the right). The diagonal marked with a dashed white line denotes the vanishing particle-to-hole transmission $t_{ph}$ in the \textit{ph-ph} scattering regime. The vertical white line in the lower right panel indicates the energy cutoff $\Lambda_\mu=34\ \mathrm{meV}$ in our simulation. The encircled (C) and (E) mark two pairs of chemical potentials supporting \textit{ph-ph} and \textit{ph-p} scattering regimes, respectively. 
        (b) Scattering amplitudes as a function of the incident angle $\theta_h^{L,+}$ for chemical potentials marked in (a). The scattering amplitudes show discontinuity at the Brewster angles denoted by $\theta_b^R$. 
        (c) Local densities of the polarization (LDOP), defined in Eq.~\eqref{eq:LDOP}, for electrons injected from a metallic lead attached in the middle of the left side of the sample, obtained numerically by solving Eq.~\eqref{eq:scattering equation1}. The interface is tilted so that the incident angle $\theta_h^{L,+}=\pi/6$. The parameters and the combination of the chemical potentials are the same as in Fig.~\ref{Fig:scattering amplitudes for angles and local densities}. Moreover, the metallic lead, marked by the grey rectangle on the left, has a parabolic dispersion with the positive effective mass $\mathcal{M}_2$ and a constant energy offset $-|\mathcal{M}_0|$ being the same as the respective parameters in the $p$- and $n$-doped regions in the rest of the system.
        }
		\label{Fig:particle-like spinor}
\end{figure}
%%%%%%%%%%%%%%%%%%%%%%%%%%%%%%%%%%%%%%%%%%%%%%%%%%
In an experimental realization, electrons entering from the metallic leads are carried by the particle- and hole-like spinors in the system. The incident state $\vec{\Phi}^{l,+}$ is thus always a superposition of spinors with both types of polarization, where the superposition depends on the details of the coupling between the inverted-band system with the leads.
While in the main text, we restrict our discussion to only incident hole-like spinors, here we show the numerical results for the incident particle-like spinors. From such initial conditions, one can obtain the other four scattering amplitudes, namely $|r_{pp}|$, $|r_{ph}|$, $|t_{ph}|$ and $|t_{pp}|$, see also Appendix~\ref{app:scattering matrix approach}.
In Fig.~\ref{Fig:particle-like spinor}(a), the analytical results for all four amplitudes are shown as a function of the chemical potentials in the left and right parts and for the incident angle $\theta_h^{+,L}=\pi/6$. We find that the particle-to-hole transmission amplitude $|t_{ph}|$ is a magnitude smaller than the other scattering amplitudes $|t_{pp}|$, $|r_{ph}|$ and $|r_{pp}|$, which suggests a negligibly small contribution to the negative refraction from the incident particle-like spinors in the whole parameter space. The corresponding termination line of the \textit{ph-ph} region $\mu_p'$ (solid white line) is given by $\mu_p'=[(4\sqrt{\mu_{R}^{2}-\mathcal{A}_c^2}+3\mathcal{M}_{0})^2+\mathcal{A}_{c}^2]^\frac{1}{2}$ for $\mu_R>\sqrt{9\mathcal{M}_{0}^2/16 +\mathcal{A}_{c}^2}$. 

Moreover, we show in Fig.~\ref{Fig:particle-like spinor}(b) that the transmission amplitude $|t_{ph}|$ remains small for all the incident angles, and only one Brewster's angle $\theta_{b}^R$ appears due to the termination of the inner branch in the $n$-doped region. Hence, we find that the incident particle-like spinors merely serve as a homogeneous background to the refocusing electron beams in the $n$-doped region, which is attributed to the incident hole-like spinors.

In Fig.~\ref{Fig:particle-like spinor}(c), a realistic experimental situation is shown with a lead made of a normal metal, i.e., with a parabolic dispersion containing only particle-like states. The LDOP is calculated numerically~\cite{Groth_2014}, and shown for two pairs of chemical potentials $\mu_L$ and $\mu_R$ lying in \textit{ph-ph} and \textit{ph-p} scattering regimes marked with (C) and (E), respectively. The negative refraction governed by the $t_{hp}$ scattering can be seen in both (C) and (E) regimes, with a difference that in (C), the effect occurs only at very large angles, i.e. $\theta_h^{L,+} > \theta_b^L$, as follows from Fig.~\ref{Fig:scattering amplitudes for angles and local densities}(a). All other transmitted waves, namely the ones that come from $t_{ph}$, $t_{hh}$ and $t_{pp}$, only give rise to the homogeneous background in the LDOP.  

%%%%%%%%%%%%%%%%%%%%%%%%%%%%%%%%%%%%%%%%%%%%%%%%%%%%%%
%%%%%%%%%%%%%%%%%%%%%%%%%%%%%%%%%%%%%%%%%%%%%%%%%%%%%%
\section{Scattering amplitudes in the asymmetric junction   \label{sec:app_asmmetric_junction}}
For the asymmetrical junction, we find that the scattering processes can be characterized by two factors (i) the momentum transfer between the scattering waves, and (ii) the cross-product between corresponding band spinors. We further find that the spinor-dependent factor remains constant along momenta lying on the same branch of the Fermi surface [see in Fig.~\ref{Fig:band spinor and bound states}(a)] -- so that, for fixed chemical potentials, the scattering amplitudes are determined solely by a momentum-dependent factor. 

Let us now calculate the scattering amplitudes for asymmetric junctions. We consider a hole-like state impinging on the interface from the left side, and study the region of chemical potentials that give rise to a \textit{ph-ph} scattering regime. Without loss of generality, we assume $0<|\mu_R|<|\mu_L|<\Lambda_\mu$, such that the $k_x$ momenta is confined to $0<|k^{L,-}_p|<|k^{R,-}_h|<|k^{R,-}_p|<|k^{L,+}_h|$. For simplicity, we denote their absolute values using $k_1=|k^{L,-}_p|$, $k_2=|k^{R,-}_h|$, $k_3=|k^{R,-}_p|$ and $k_4=|k^{L,+}_h|$, and rewrite the scattering equations in Eq.~\eqref{eq: continuity condition} as
{\footnotesize
\begin{align} \label{eq:scattering equation asymm}
     \vec{\Psi}^{L,+}_h+\tilde{r}_{hh}\vec{\Psi}^{L,-}_h+\tilde{r}_{hp}\vec{\Psi}^{L,-}_p &= \tilde{t}_{hp}\vec{\Psi}^{R,-}_p+\tilde{t}_{hh}\vec{\Psi}^{R,-}_h \nonumber \\
     k_4 \left( -\vec{\Psi}^{L,+}_h + \tilde{r}_{hh}\vec{\Psi}^{L,-}_h \right)- k_1\tilde{r}_{hp}\vec{\Psi}^{L,-}_p &= 
     k_3\tilde{t}_{hp}\vec{\Psi}^{R,-}_p-k_2\tilde{t}_{hh}\vec{\Psi}^{R,-}_h.
\end{align}
}
By solving the above equations, we obtain the scattering amplitudes in the following form
\begin{align}\label{eq: amplitudes asymm}
    \tilde{t}_{hp}&=\frac{k_4-k_1}{k_3+k_1}\frac{\vec{\Psi}^{L,+}_{h}\cross\vec{\Psi}^{L,-}_{h}}{\vec{\Psi}^{L,-}_{h}\cross\vec{\Psi}^{R,-}_{p}}-\frac{k_2-k_1}{k_3+k_1}\frac{\vec{\Psi}^{R,-}_{h}\cross\vec{\Psi}^{L,-}_{h}}{\vec{\Psi}^{L,-}_{h}\cross\vec{\Psi}^{R,-}_{p}}\tilde{t}_{hh},\nonumber\\    
    \tilde{t}_{hh}&=\frac{2k_4}{k_4+k_2}\frac{\vec{\Psi}^{L,+}_{h}\cross\vec{\Psi}^{L,-}_{p}}{\vec{\Psi}^{R,-}_{h}\cross\vec{\Psi}^{L,-}_{p}}-\frac{k_4-k_3}{k_4+k_2}\frac{\vec{\Psi}^{R,-}_{p}\cross\vec{\Psi}^{L,-}_{p}}{\vec{\Psi}^{R,-}_{h}\cross\vec{\Psi}^{L,-}_{p}}\tilde{t}_{hp},\nonumber\\
    \tilde{r}_{hp}&=\frac{k_3+k_4}{k_3+k_1}\frac{\vec{\Psi}^{L,+}_{h}\cross\vec{\Psi}^{R,-}_{h}}{\vec{\Psi}^{R,-}_{h}\cross\vec{\Psi}^{L,-}_{p}}-\frac{k_4-k_3}{k_3+k_1}\frac{\vec{\Psi}^{L,-}_{h}\cross\vec{\Psi}^{R,-}_{h}}{\vec{\Psi}^{R,-}_{h}\cross\vec{\Psi}^{L,-}_{p}}\tilde{r}_{hh},\nonumber\\
    \tilde{r}_{hh}&=\frac{k_4-k_2}{k_4+k_2}\frac{\vec{\Psi}^{L,+}_{h}\cross\vec{\Psi}^{R,-}_{p}}{\vec{\Psi}^{L,-}_{h}\cross\vec{\Psi}^{R,-}_{p}}-\frac{k_2-k_1}{k_4+k_2}\frac{\vec{\Psi}^{L,-}_{p}\cross\vec{\Psi}^{R,-}_{p}}{\vec{\Psi}^{L,-}_{h}\cross\vec{\Psi}^{R,-}_{p}}\tilde{r}_{hp} \,.
\end{align}
The scattering amplitudes written above can be solved iteratively. Note that in the symmetric case, where $k_1=k_2$ and $k_3=k_4$, the expressions in Eq.~\eqref{eq: amplitudes symm} are recovered.

%%%%%%%%%%%%%%%%%%%%%%%%%%%%%%%%%%%%%%%%%%%%%%%%%%%%%%%%%%
%%%%%%%%%%%%%%%%%%%%%%%%%%%%%%%%%%%%%%%%%%%%%%%%%%%%%%%%%%
\section{Bound states at the interface  \label{sec:app_bound_states}}
%%%%%%%%%%%%%%%%%%%%%%%%%%%%%%%%%%%%%%%%%%%%%
\begin{figure}
		\centering
		\includegraphics[scale=1]{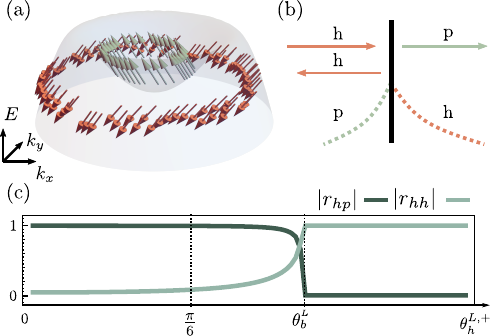}
		\caption{(a) Band spinors located at the Fermi surface in the $p$-doped region. Arrows aligning downwards at the outer branch represent the hole-like spinors (red), while the upwards-aligning arrows located at the inner branch represent the particle-like spinors (green). (b) Sketch of the scattering processes with bound states in the \textit{h-p} regime. (c) Scattering amplitudes as a function of the incident angle $\theta_h^{L,+}$ for chemical potentials $\mu_L=18\ \mathrm{meV}$ and $\mu_R=0\ \mathrm{meV}$.}
		%A hole-like state impinging on the interface can be either reflected to hole-like state or transmitted to particle-like state. At the same time, the imaginary solutions of $k_x$ correspond to the exponentially decaying bound states located at the interface. Here we show the hole-like bound state from the $p$-doped region and the particle-like bound state from the $n$-doped region.}
		\label{Fig:band spinor and bound states}
\end{figure}
%%%%%%%%%%%%%%%%%%%%%%%%%%%%%%%%%%%%%%%%%%%%%%%%%%
In Sec.~\ref{subsec:scattering amplitudes}, the presence of the bound states in the \textit{h-p} scattering regime alters the scattering amplitudes, which then reads as
\begin{align}   \label{eq:BS amplitudes}
    \tilde{t}_{hh}^{BS}&=\frac{2k_{i}}{k_-}\frac{\vec{\Psi}^{L,+}_{h}\cross\vec{\Psi}^{L,-}_{p}}{\vec{\Psi}^{R,-}_{h}\cross\vec{\Psi}^{L,-}_{p}} \, , \nonumber\\
    \tilde{r}_{hp}^{BS}&=\frac{2k_{i}}{k_+}\frac{\vec{\Psi}^{L,+}_{h}\cross\vec{\Psi}^{R,-}_{h}}{\vec{\Psi}^{R,-}_{h}\cross\vec{\Psi}^{L,-}_{p}} \, , \nonumber\\
    \tilde{t}_{hp}^{BS}&=\frac{k_{-}}{k_+}\frac{\vec{\Psi}^{L,+}_{h}\cross\vec{\Psi}^{L,-}_{h}}{\vec{\Psi}^{L,-}_{h}\cross\vec{\Psi}^{R,-}_{p}}-\frac{2k^{BS}}{k_+}\tilde{t}_{hh}^{BS}\frac{\vec{\Psi}^{R,-}_{h}\cross\vec{\Psi}^{L,-}_{h}}{\vec{\Psi}^{R,-}_{p}\cross\vec{\Psi}^{L,-}_{h}} \, , \nonumber\\
    \tilde{r}_{hh}^{BS}&=\frac{k_{-}}{k_+}\frac{\vec{\Psi}^{L,+}_{h}\cross\vec{\Psi}^{R,-}_{p}}{\vec{\Psi}^{L,-}_{h}\cross\vec{\Psi}^{R,-}_{p}}-\frac{2k^{BS}}{k_+}\tilde{t}_{hh}^{BS}\frac{\vec{\Psi}^{R,-}_{h}\cross\vec{\Psi}^{R,-}_{p}}{\vec{\Psi}^{R,-}_{p}\cross\vec{\Psi}^{L,-}_{h}} \, ,
\end{align}
where $k_{\pm}=|k^{L,+}_h|\pm k^{BS}$ and $k^{BS}=i|\mathrm{Im}(k^{L,-}_p)|$.

The selection of the incoming states stems from the fact that Eq.~\eqref{eq:kx} yields two solutions of $k_x$ for each branch in all scattering regimes. In the \textit{ph-ph} regime, all solutions are real and each one of them represents a propagating state. On the other hand, in the \textit{h-p}, \textit{ph-p} and \textit{h-ph} regimes, the inner branch of the Fermi surface vanishes for given chemical potentials and incident $k_y$, and therefore, the corresponding $k_x$ solutions are imaginary and represent exponentially decaying solutions located at the interface between $p$ and $n$ region, i.e., bound states, see Fig.~\ref{Fig:band spinor and bound states}(b). Their localization length $\xi$ is given by the reciprocal of the imaginary part of $k_x$, i.e., $\xi=2\pi/\mathrm{Im}(k_x)$. Equations~\eqref{eq:BS amplitudes} show that although the bound states carry no propagating charges, they are a crucial correction term for the scattering amplitudes and are thus needed to be included in the outgoing states. Furthermore, we find that the aforementioned expressions of the scattering amplitudes in Eq.~\eqref{eq: amplitudes asymm} are valid even in the \textit{h-p}, \textit{ph-p} and \textit{h-ph} regimes once the imaginary part of $k_x$ is included, namely by defining $k_2=-i|\mathrm{Im}(k^{R,-}_h)|$ and $k_1=i|\mathrm{Im}(k^{L,-}_p)|$. Note that these definitions further allow us to calculate the scattering amplitudes when one of the regions is fully gapped, i.e., when only bound states exist in one of the two regions. For instance, in the case where the Fermi surface of the $n$-doped region lies in the main gap.  We show in Fig.~\ref{Fig:band spinor and bound states}(c) the scattering amplitudes $r_{hp}$ and $r_{hh}$ for a hole-like spinor impinging from the $p$-doped region. The dominant reflection changes from $r_{hp}$ to $r_{hh}$ as the incident angle increases, which inspired us to build the polarization-controllable mirror presented in Sec.~\ref{subsec: Mirror}.
%%%%%%%%%%%%%%%%%%%%%%%%%%%%%%%%%%%%%%%%%%%%%%%%%%%%%%%%%%%
%%%%%%%%%%%%%%%%%%%%%%%%%%%%%%%%%%%%%%%%%%%%%%%%%%%%%%%%%%%%%%
\newpage

\end{document}